\documentclass[preprint2]{aastex}

\usepackage{graphicx}
\usepackage{amsfonts}
\usepackage{times}

\newcommand{\vk}{\mbox{\boldmath $ k $}}

\newcommand{\vB}{\mbox{\boldmath $ B $}}
\newcommand{\vrr}{\mbox{\boldmath $ r $}}
\newcommand{\vb}{\mbox{\boldmath $ b $}}
\newcommand{\vbo}{\vb^0}

\newcommand{\vU}{\mbox{\boldmath $ U $}}
\newcommand{\vu}{\mbox{\boldmath $ u $}}
\newcommand{\vuo}{\vu^0}
\newcommand{\vuoe}{\vu^{0,e}}
\newcommand{\oB}{\overline{\vB}}

\newcommand{\oU}{\overline{\vU}}
\newcommand{\vOmega}{\mbox{\boldmath $ \Omega $}}
\newcommand{\btimes}{\mbox{\boldmath $ \times $}}

\shorttitle{Waves, Coriolis force and the dynamo effect}
\shortauthors{Mininni, G\'omez, \& Mahajan}

\begin{document}

\title{ WAVES, CORIOLIS FORCE AND THE DYNAMO EFFECT}

\author{Swadesh M. Mahajan}
\affil{Institute for Fusion Studies, The University of Texas,
       Austin, Texas 78712, USA.}
\and
\author{Pablo D. Mininni\altaffilmark{1}}
\affil{Advanced Study Program, National Center for
       Atmospheric Research, P.O.Box 3000, Boulder CO 80307, USA.}
\and
\author{Daniel O. G\'omez\altaffilmark{2}}
\affil{Departamento de F\'\i sica,
       Facultad de Ciencias Exactas y Naturales,
       Universidad de Buenos Aires, \\
       Ciudad Universitaria, 1428 Buenos Aires, Argentina.}
       \altaffiltext{1}{also at Departamento de F\'\i sica,
       Facultad de Ciencias Exactas y Naturales, Universidad de Buenos Aires,
       Ciudad Universitaria, 1428 Buenos Aires, Argentina.}
       \altaffiltext{2}{also at Instituto de Astronom\'\i a y F\'\i sica del
              Espacio, Ciudad Universitaria, 1428 Buenos Aires, Argentina.}
        \email{dgomez@df.uba.ar}

\begin{abstract}

Dynamo activity caused by waves in a rotating magneto-plasma 
is investigated. In astrophysical environments such as accretion disks
and at sufficiently small spatial scales, the Hall effect is likely 
to play an important role. It is shown that  a combination of the 
Coriolis force 
and Hall effect can produce a finite $\alpha$-effect by generating net 
helicity  in the small scales. The shear/ion-cyclotron  normal mode of the 
Hall plasma is the dominant contributor to the dynamo action for short 
scale motions.
 
 \end{abstract}

\section{INTRODUCTION}

In astrophysical objects, large scale magnetic fields are thought to be
generated by helical turbulence (the so-called $\alpha$-effect) and differential 
rotation (the $\Omega$-effect) (see \citet{Meneguzzi,Brandenburg}). 
However, note that a large scale magnetic field
can in some cases be generated without fully helical turbulence or a net
$\alpha$-effect. Within the mean field approximation, there are physical 
effects which contribute to the mean electromotive force even when the
$\alpha$ coefficient is zero. A shear turbulent flow \citep{Urpin}, 
the $\omega\times j$ term \citep{Radler69, Geppert}, or magnetic instabilities 
in a stably stratified atmosphere \citep{Spruit}, are some of the examples 
reported in the literature.

Helicity is naturally imparted to a rotating fluid by the Coriolis force 
\citep{Moffat1,Moffat2,Moffat3,Moffat4}. However, at sufficiently small scales, 
where the Rossby number 
\begin{equation}
R_S = \frac{U_0}{2 L_0 \Omega}
\label{Rossby}
\end{equation}
is larger than unity ($ U_0 $ and $ L_0 $ are characteristic velocities 
and lengths, and $ \Omega $ is the rotation rate), the Coriolis force and the 
resulting induced helicity might become negligibly small. Nonetheless, it is 
worth noting that in helical turbulent flows, the kinetic helicity develops a direct 
cascade along with the energy \citep{Eyink,Gom2004}. Therefore, some small scale 
turbulent flows can still be helical, even though the source of helicity remains at 
much larger spatial scales.

The Hall effect introduces a definite handedness or helicity on small scale fluid 
motions (\citet{Wardle}, \citet{Balbus}), since the mirror symmetry in the induction 
equation is broken (see Eqn~(\ref{induc}) below). Therefore, one  should expect a net 
$\alpha$-effect  in a Hall plasma. The Hall effect becomes relevant whenever the Hall 
length scale 
\begin{equation}
\lambda = \frac{c}{\omega_{pi}}\ \frac{U_A}{U_0} \; .
\label{LHall}
\end{equation}
is larger than the dissipation scale, a category to which several objects 
of astrophysical interest belong \citep{MGM1,MGM3}. Here, $ U_A $ is the 
characteristic Alfv\'enic speed, $c$ is the speed of light, and $\omega_{pi}$ is 
the ion plasma frequency.

In \S 2 we write down the Hall MHD equations. The normal modes  sustained 
in this system are derived and listed in \S 3. In \S 4 we briefly summarize the role of 
normal mode fluctuations on MHD dynamos, and in \S 5 this concept is extended 
to Hall MHD. In \S 6 we show the effect of rotation on the $\alpha$-effect. The 
main results of the present work are summarized in \S 7.

\section{THE HALL-MHD SYSTEM}

The dynamics of ideal and incompressible fully ionized plasmas in a rotating 
frame, is described by the induction equation (modified by the addition of 
the Hall current) and the Navier-Stokes equation,
\begin{eqnarray}
\frac{\partial \vB}{\partial t} & = & \nabla \btimes \left[ \left( \vU
     - \lambda \nabla \btimes \vB \right) \btimes \vB \right] 
\label{induc} \\
\frac{\partial \vU}{\partial t} & = & - \left( \vU \cdot \nabla \right) \vU  
     - 2 \, \vOmega \btimes \vU + \left( \nabla \btimes \vB \right) \btimes 
     \vB - {}
     \nonumber\\
 & & {} - \nabla \left( P - \left| \vOmega \btimes \vrr \right|^2 \right) \; ,
\label{NS}
\end{eqnarray}
with the constraints 
\begin{equation}
\nabla \cdot \vU = \nabla \cdot \vB = 0 \; .
\label{div}
\end{equation}
These equations are known as the Hall-MHD equations. The magnetic field is 
expressed in velocity units, i.e. $\vB = {\cal B}[Gauss] (4\pi\rho)^{-1/2}$, 
where $\rho$ is the constant mass density. The quantity $P$ is the 
gas pressure divided by the constant mass density.  We define a dimensionless 
number $ \epsilon $ to measure the relative strength of the Hall effect,
\begin{equation}
\epsilon = \frac{\lambda}{L_0} \; ,
\end{equation}
where $ L_0 $ is the characteristic length scale of the system.  
If we choose $U_0 = U_A $ as the characteristic velocity, $\lambda$ reduces 
to the ion skin depth. Note that equation (\ref{LHall}) is valid for a fully 
ionized plasma.

\section{WAVES IN HALL-MHD}

We will study waves in the ideal and incompressible Hall-MHD system in a 
rotating frame. To linearize the Hall-MHD equations around a static and 
uniform magnetic field $\vB_0$, we write
\begin{eqnarray}
\vB & = & \vB_0 + \vb \\
\vU & = & \vu \; ,
\end{eqnarray}
where $ \vu ,\vb \sim \exp(i \vk \cdot \vrr - i \omega t)$. These substitutions
convert (\ref{induc}) and (\ref{NS}) into the closed set :
\begin{eqnarray}
-\omega \vb & = & \vk \btimes \left[ \left( \vu - i \lambda \vk \btimes 
     \vb \right) \btimes \vB_0 \right] \\
-\omega \vu & = & 2i \, \vOmega \btimes \vu + \left( \vk \btimes \vb 
     \right) \btimes \vB_0 - \vk P_{tot} \; ,
\label{linNSwithP}
\end{eqnarray}
where the total effective pressure  $ P_{tot} = P - | \vOmega \btimes \vrr |^2 $. 

The elimination of $ P_{tot} $ in incompressible flows is arranged by projecting onto the 
plane perpendicular to $ \vk $ . One finally obtains
\begin{eqnarray}
-\omega \vb & = & \left( \vk \cdot \vB_0 \right) \left( \vu - i \lambda 
     \vk \btimes \vb \right) 
\label{lininduc} \\
-\omega \vu & = & \mathbb{P} \left[ 2i \, \vOmega \btimes \vu + \left( 
     \vk \btimes \vb \right) \btimes \vB_0 \right] \; ,
\label{linNS}
\end{eqnarray}
where $ \mathbb{P}_{ij} = \delta_{ij}-k_i k_j k^{-2} $ is the projector 
operator.

Without loss of generality we choose $\vk$ in the $z$-direction, 
i.e. $\vk = k\hat{z} $, while $ \vOmega $ and $ \vB_0 $ can be oriented 
arbitrarily.  Equations (\ref{lininduc}) and (\ref{linNS}), then, 
reduce to
\begin{equation}
\left( \omega \mathbb{I} + i \lambda k^2 B_z \mathbb{A} \right) \vb 
      =  - k B_z \mathbb{I} \vu 
\label{lininduc2}
\end{equation}
\begin{equation}
\left( \omega \mathbb{I} - 2i \Omega_z \mathbb{A} \right) \vu 
      =  - k B_z \mathbb{I} \vb \; .
\label{linNS2}
\end{equation}
Here,
\begin{equation}
\mathbb{I} = \left( \begin{array}{rrr}
1 &  0 &  0 \\
0 &  1 &  0 \\
0 &  0 &  0 \end{array} \right) \; ,
\end{equation}
and
\begin{equation}
\mathbb{A} = \left( \begin{array}{rrr}
0  &  1 &  0 \\
-1 &  0 &  0 \\
0  &  0 &  0 \end{array} \right) \; .
\end{equation}

Note that only the $z$-components of $\vOmega$ and $\vB_0$ are relevant 
(i.e. the components in the direction of the vector $\vk$). The orthonormal 
base for the antisymmetric operator $ \mathbb{A} $ is given by
\begin{equation}
\left| \pm \right> = \frac{1}{\sqrt{2}} \left( \begin{array}{r}
1  \\  \pm i \\ 0 \end{array} \right) \; ,
\label{ket}
\end{equation}
and satisfies $ \mathbb{A} \left| \pm \right> = \pm i \left| \pm \right> $. 

The dispersion relation can be obtained from Equations (\ref{lininduc2}) and 
(\ref{linNS2}),

\begin{equation}
\omega^2 - k^2 \left( B_z^2 + 2 \lambda \Omega_z B_z 
     \right) + \sigma \omega \left( 2 \Omega_z - \lambda 
     k^2 B_z \right) = 0 \; ,
\label{disprel}
\end{equation}
where $ \sigma \equiv \pm 1 $ and the eigenvectors can be written as 
$\left| \sigma \right>$.

This dispersion relationship is quite general and includes several well 
known   waves of a magnetized plasma in the corresponding asymptotic limits. When both 
$ \Omega_z = 0 $ (no rotation) and $ \lambda = 0 $ (negligible Hall current, 
i.e. the MHD limit) we obtain \ \ $\omega^2 - k^2 B_z^2 = 0 $ which 
corresponds to Alfv\'en waves. In the incompressible limit, the shear and the 
compressional waves are degenerate and indistinguishable.
When $ \lambda = 0 $, equation (\ref{disprel}) 
reduces to $ \omega^2 + 2 \sigma \omega \Omega_z - k^2 B_z^2 =0 $, and 
we obtain the inertial waves first described in the absence of an external 
magnetic field $ B_z $ by \citet{Moffat1,Moffat2,Moffat3}. Finally, when 
$ \Omega_z = 0 $ we obtain two branches of circularly polarized waves. The 
right hand polarized branch, corresponds to whistlers, with a frequency 
growing like $\omega\simeq k^2$ at large wavenumbers. These waves 
are the high k limit of the compressional branch. The left handed 
branch is the standard shear wave  whose frequency approaches 
the ion-cyclotron frequency asymptotically.

\begin{figure}
\plotone{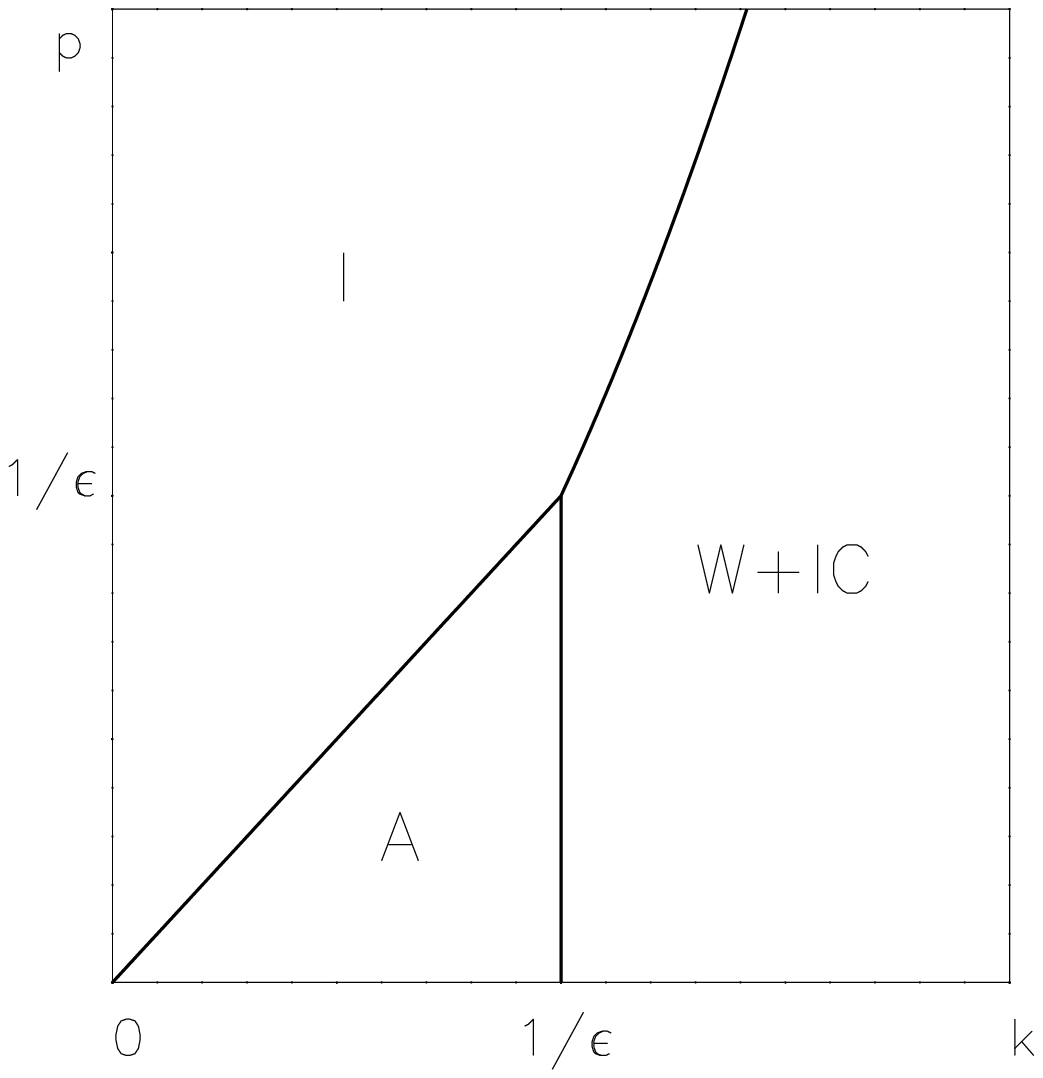}
\caption{Regions dominated by: whistlers (W) and ion-cyclotron (IC) waves, 
inertial (I), and Alfv\'en (A) waves. The horizontal axis is the 
wavenumber in units of $1/L_0$. The vertical axis corresponds to 
$p = 2 \Omega_z L_0 / B_z$.
\label{dispfig}}
\end{figure}

Figure~\ref{dispfig} shows these asymptotic cases in detail. The general 
dispersion relationship given in equation (\ref{disprel}), can be cast in 
dimensionless units using $L_0$ and $B_z = U_A$ as typical longitude and velocity, 
\begin{equation}
\omega^2 - k^2 (1 + p \epsilon) + \sigma\omega (p - \epsilon k^2) = 0 \; ,
\label{displess}
\end{equation}
where
\begin{equation}
p = \frac{2 \Omega_z L_0}{B_z} \; ,
\label{pe}
\end{equation}
can be interpreted as the inverse of the Rossby number given in Eq. (\ref{Rossby}).
The MHD limit in Figure \ref{dispfig} corresponds to a small neighborhood around the origin, 
where there is only a transition between inertial waves and Alfv\'en waves.

\begin{figure}
\plotone{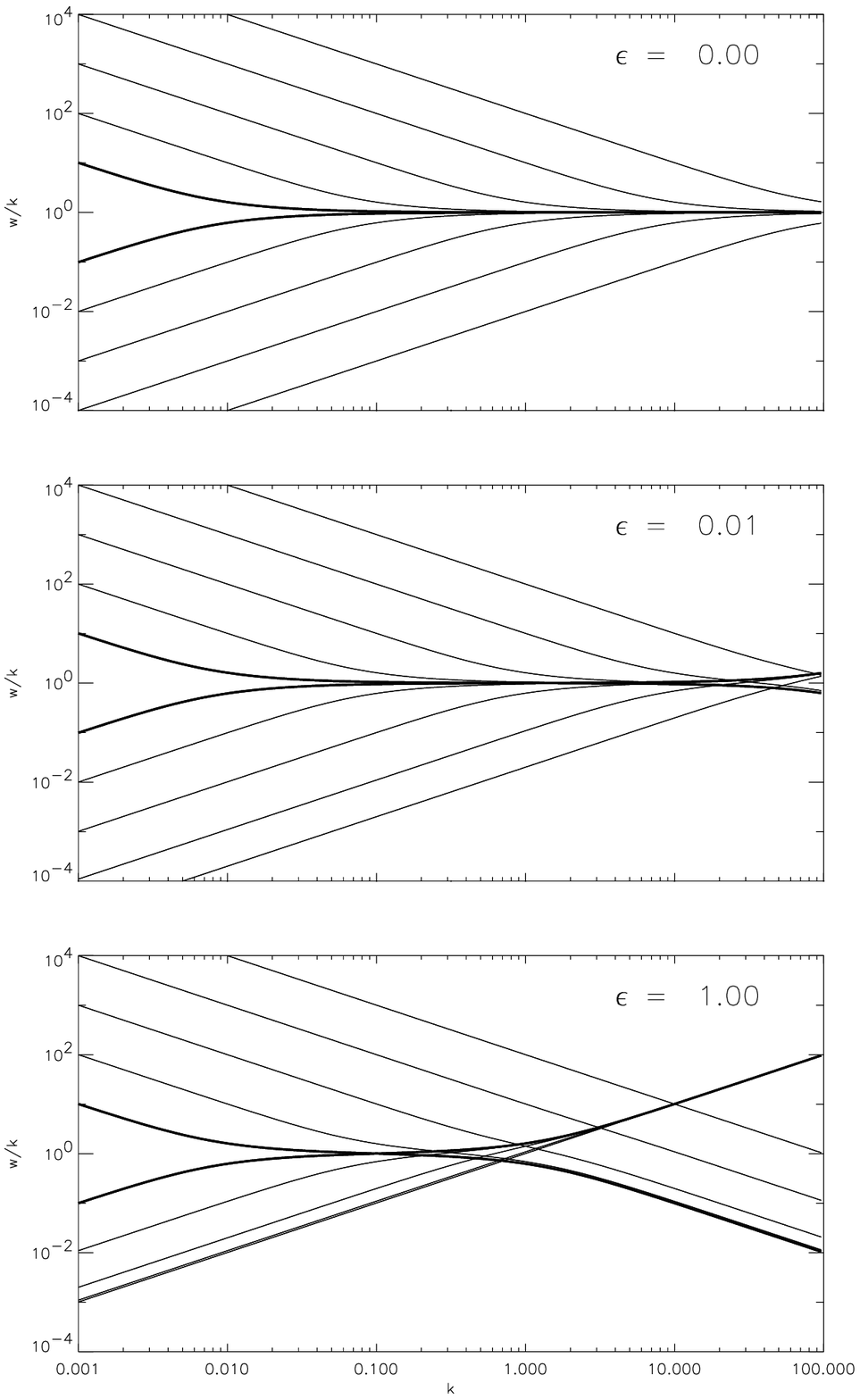}
\caption{Phase speed vs. wavenumber for the normal modes arising from the normalized 
dispersion relationship given in equation (\ref{displess}). Each frame corresponds to a 
different value of $\epsilon$, while different traces correspond to $p= 0.01$ (thick) and 
$p = 0.1, 1, 10, 100$.
\label{v_phase}}
\end{figure}

Figure \ref{v_phase} shows the phase speed as a function of wavenumber for the two positive 
branches given by equation (\ref{displess}). There are also two other identical branches with 
negative frequencies (not shown). These waves behave approximately as Alfv\'en waves only in 
the wavenumber regions where they are non-dispersive, i.e. when the curves become horizontal.

\section{MHD WAVES AND THE DYNAMO EFFECT}

Before embarking on the Hall-MHD case, it is instructive to recall
some previous results relating to the dynamo process induced by MHD waves ( readily  
obtained from our expressions). In the limit $ \vOmega \to 0 $ and 
$ \lambda \to 0 $, Alfv\'en waves are known to quench the 
$ \alpha $-effect \citep{Gruzinov}. The MHD $\alpha$-coefficient 
\citep{Pouquet,Blackman} is given by 
\begin{equation}
\alpha = \frac{\tau}{3} \left( -\left< \vu \cdot \nabla \btimes \vu \right> 
    + \left< \vb \cdot \nabla \btimes \vb \right> \right ) \; ,
\label{alphaMHD}
\end{equation}
where $ \vu $ and $ \vb $ are respectively small scale velocity and magnetic 
fields, unaffected by the presence of a large scale magnetic field (see details in 
\citet{Blackman}), and the coefficient $\tau$ is a typical correlation time for the 
turbulent small scale motions. A pure Alfvenic state satisfies
$ \vu = \pm \vb $ and therefore $ \alpha = 0 $. This is to
be expected, since all nonlinear terms cancel exactly for Alvenic states, 
and therefore no transportation coefficients can arise in that case. On the other 
hand, when $ \vOmega \neq 0 $ the Coriolis force is expected to inject helicity to
the fluid, and therefore a net $ \alpha $-effect arises \citep{Moffat4}. 
A detailed discussion can also be found in \citet{Moffat1,Moffat2,Moffat3}

\section{HALL-MHD NORMAL MODES AND THE $\alpha$-EFFECT}

 The first studies on the impact of Hall currents on dynamo action 
\citep{Helmis1,Helmis2} were carried out using mean field theory and the 
first-order smoothing approximation \citep{Krause}. Helmis obtained decreasing 
dynamo action as the strength of the Hall terms increased. Recently, the relevance 
of the Hall effect on dynamo activity was confirmed experimentally by \citet{Ding}.

In \citet{MGM1} it was shown that the expression for the $ \alpha $-effect 
in the presence of the Hall effect is modified according to
\begin{eqnarray}
\alpha & = & \frac{\tau}{3} \left( - \left< \vu^e \cdot \nabla \btimes
     \vu^e \right> + \left< \vb \cdot \nabla \btimes \vb \right> - 
     \right. {}
     \nonumber\\
 & & {} \left. - \lambda \left< \vb \cdot \nabla \btimes \nabla \btimes 
     \vu^e \right> \right) \; ,
\label{alpha}
\end{eqnarray}
where $ \vu^e \equiv \vu - \lambda \nabla \btimes \vb $ is the small 
scale electron flow velocity. This general expression differs from the 
MHD result (equation (\ref{alphaMHD})) in two ways: it replaces the 
kinetic helicity by the helicity of the electron flow, and it contains an 
extra term due to the Hall current in the microscale. A nontrivial 
consequence of the latter is that, while the original $ \alpha $ 
coefficient of \citet{Pouquet} is zero for a pure Alfv\'enic state 
$ \vu = \sigma \vb $ \citep{Gruzinov}, the one corresponding to 
equation (\ref{alpha}) is not. In \citet{MGM3} the impact of the Hall 
effect in helical turbulent dynamos was studied in direct numerical 
simulations. As mentioned in the previous section, the required helicity 
is naturally introduced in a fluid in a rotating body by the Coriolis 
force.

Two questions arise.  The first one is better posed and answered for a non-rotating plasma.
We just showed that the $\alpha$ derived in MHD goes to zero for the pure
Alfv\'enic state , which is an eigen-state of MHD. We also claimed  that for the same pure eigen-state 
of MHD, the $\alpha$ derived in Hall MHD does not vanish. Surely that is an unwarranted mixing of different worlds. What we should, instead, calculate is the value of $\alpha$ derived in Hall MHD for the corresponding normal modes of Hall MHD. This task is performed at the end of this section.

The  second question concerns rotation and the  
Coriolis force which is known to   inject helicity at large scales (see equation 
[\ref{Rossby}] and Figure \ref{dispfig}). At small scales (large $k$), however, the 
Coriolis force is not considered to be relevant (motions are expected to be essentially non-helical) posing a serious restriction on the generation of  magnetic fields in astrophysics. 
Can we find a source of helicity acting at small scales, and will this source have much to do with rotation? 

We show that 
the Hall effect naturally introduces helicity at small scales, which is 
precisely the region where this effect is stronger (in agreement with \citet{Wardle} 
and also \citet{Balbus}). However, no net $\alpha$-effect is generated by these 
microscale motions, unless there is also a net rotation of the system; a combination
of rotation and Hall effect is needed for dynamo action.

As is shown in equation (\ref{ket}), the general solutions of the 
linearized Hall-MHD equations in a rotating frame are right-handed or 
left-handed polarized waves. To investigate
the effect of Hall currents at small scales, we first  concentrate on a non-rotating plasma,
 $ \Omega_z = 0 $. For large k, the 
dispersion relationship (\ref{disprel}) reduces to 
\begin{equation}
\omega^2 - k^2 B_z^2 - \sigma \lambda \omega k^2 B_z = 0 \; ,
\label{dispnorot}
\end{equation}
which coupled to Eq.(\ref{lininduc}), yields
\begin{equation}
\vb = - \frac{k B_z}{\omega} \vu_e \; .
\label{ueandb}
\end{equation}
Inserting Eq. (\ref{ueandb}) into equation (\ref{alpha}), and invoking 
that the electron vorticity is $ \nabla \btimes \vu_e = \sigma k \vu_e $, 
we obtain 
\begin{eqnarray}
\alpha  = - \frac{\tau}{3} \left( \frac{\omega^2 - k^2 B_z^2 - \sigma \lambda
\omega k^2 B_z}{\omega^2}\right) \left< \vu^e \cdot \nabla \btimes \vu^e 
\right>
\label{alpha2}
\end{eqnarray}
which, according to the dispersion relation Eq. (\ref{dispnorot}), 
corresponds to $\alpha = 0$. This result of a zero $\alpha$-effect for Hall MHD normal modes
is an expected and  natural extension of the similar result  obtained for Alfvenic states in MHD. 
However, we must bear in mind that these results are derived for non-rotating 
systems.  

\section{CORIOLIS FORCE AND THE $\alpha$-EFFECT}

In rotating objects, the closure calculation leading to either equation (\ref{alpha}) (in 
the MHD limit) or equation (\ref{alphaMHD}) needs to be revised, since the Coriolis force 
(see equation (\ref{NS})) was not included in those calculations. The starting point 
is the so-called ``{\it reduced smoothing approximation}" (RSA) proposed by \citet{Blackman} 
(see also \citet{MGM2} for a derivation which includes the Hall effect).
Following RSA, we decompose the magnetic and velocity fields as 

\begin{equation}
\vB = \oB + \vb + \vbo
\label{expansion1}
\end{equation}

\begin{equation}
\vU = \oU + \vu + \vuo
\label{expansion2}
\end{equation}
where the overbar denotes spatially or statistically averaged large-scale
perturbations. The small scale fields $\vbo$, $\vuo$ are solutions of Eqs. 
(\ref{induc})-(\ref{NS}) in the absence of large scale fields, and $\vb$, $\vu$ are  
anisotropic corrections to the small scale fields, caused by the presence of $\oB$, $\oU$.
The net effect of small scale fluctuations on the large scale dynamics, is given by 
an electromotive force 
\begin{equation}
{\cal E} =  \left< \vuoe \times \vb + \vu^e \times \vbo  \right>
\label{EMF}
\end{equation}
acting on equation (\ref{NS}),  where $ \vu^e = \vu - \lambda\nabla\times\vb $ and 
$ \vuoe = \vu^0 - \lambda\nabla\times\vb^0 $.

From equations (\ref{induc})-(\ref{NS}) for the evolution of small scale fields (assuming 
that $\oU \approx 0$ in an appropriate frame of reference, and under the 
reduced smoothing approximation \citep{Blackman}), we obtain
\begin{equation}
\partial_t \vb \simeq \left( \oB\cdot\nabla\right) \vuoe \; ,
\label{Bmicro}
\end{equation}
\begin{equation}
\left[\partial_t + 2 \mathbb{I}\cdot\vOmega\times \right]\vu \simeq \left( \oB \cdot \nabla \right) 
 \vbo \; .
\label{Umicro}
\end{equation}

The time derivative indicated in equations (\ref{Bmicro})-(\ref{Umicro}) is usually approximated 
by $\partial_t \approx 1/\tau$, where $\tau$ is a correlation time for the microscale motions.
The operator on the left-hand side of equation (\ref{Umicro}) becomes,
\begin{equation}
\mathbb{T}^{-1} = \frac{1}{\tau} + 2 \mathbb{I}\cdot\vOmega\times 
\label{invtau}
\end{equation}
whose inverse is
\begin{equation}
\mathbb{T} = \frac{\tau}{1 + (2\Omega_z\tau)^2} \left( \mathbb{I} + 2\Omega_z\tau \mathbb{A} \right) \; ,
\label{tau}
\end{equation}

Replacing equations (\ref{Bmicro})-(\ref{Umicro}) on equation (\ref{EMF}) and using equation (\ref{tau}), 
we obtain
\begin{eqnarray}
{\cal E}&=&\frac{\tau\sigma k B_z}{\omega^2}\left(-\omega^2 + 
 \frac{k^2 B_z^2}{1 + (2\Omega_z\tau)^2} + {} \right.
 \nonumber\\
 & & {} + \left. \lambda\sigma k^2 B_z\omega \right) \left<|\vuoe|^2\right> \; .
\label{EMF2}
\end{eqnarray}

The expression for the electromotive force given in equation (\ref{EMF2}) corresponds to an 
anisotropic tensor
\begin{eqnarray}
\alpha_{ij}&=&- \frac{\sigma k \tau}{\omega^2} \left< |\vuoe|^2\right> 
   \left[2\Omega_z (\lambda k^2 B_z - 
   \sigma\omega ) + {} \right.
   \nonumber\\
   & & {} \left. + k^2 B_z^2 \left(\frac{(2\Omega_z\tau)^2}{1 + 
   (2\Omega_z\tau)^2}\right) 
   \right] \frac{k_i k_j}{k^2}
\label{alphatensor}
\end{eqnarray}
which for small scale fluctuations given by a pure mode of wavenumber $\vk$, produce an electromotive 
force $\cal E$ parallel to $\vk$, regardless of the orientation of $\oB$ and $\vOmega$. It seems 
reasonable to assume that the correlation time $\tau$ is much smaller than the rotation period.
Therefore, in the asymptotic limit $\Omega\tau \ll 1$,
\begin{equation}
\alpha_{ij} = - \frac{2\Omega_z\tau k}{\omega} \left< |\vuoe|^2\right> \left[ \frac{\lambda\sigma k^2 
        B_z}{\omega} - 1 \right] \frac{k_i k_j}{k^2} .
\label{alphatensor2}
\end{equation}
Note that when a rotation field is present, $\alpha_{ij}$ is in general non-zero. The rotation field 
therefore provides a source of energy that is responsible for the net $\alpha$-effect. Equation 
(\ref{alphatensor2}) confirms that for non-rotating objects, the net alpha effect generated by a 
background of small scale normal modes is exactly zero. Notwithstanding, sources of 
kinetic helicity other than rotation have been considered \citep{MGM2,MGM3}, to assess 
their efficiency in driving large scale dynamos.

For large wavenumbers, the dispersion relationship given by equation (\ref{disprel}) 
has two limits:  one is the so called compressional/whistler branch   $\omega \approx \pm\lambda k^2 B_z$,  while the other is the shear/cyclotron branch  with  $\omega \approx \pm\ B_z/\lambda$. For the former, the $\alpha$-effect is asymptotically small. For the shear/cyclotron branch, however,
$\alpha_{ij} \propto k^3$ at large wavenumbers implying strong dynamo action. Therefore, once 
the rotation breaks the mirror symmetry, the shear/ion-cyclotron modes ( which failed to produce 
dynamo action without rotation) are, indeed,  able to provide a net $\alpha$-effect at 
microscopic scales. We believe that this is a very important result for turbulent dynamo theories. In Figure 3 we show the trace of the tensor $\alpha_{ij}$ 
as a function of $k$, for different values of $p$ and $\epsilon$. Note that
in the MHD case $\alpha_{ii}$ drops to zero at small scales (large $k$),
while in the Hall-MHD case it does not.

\begin{figure}
\plotone{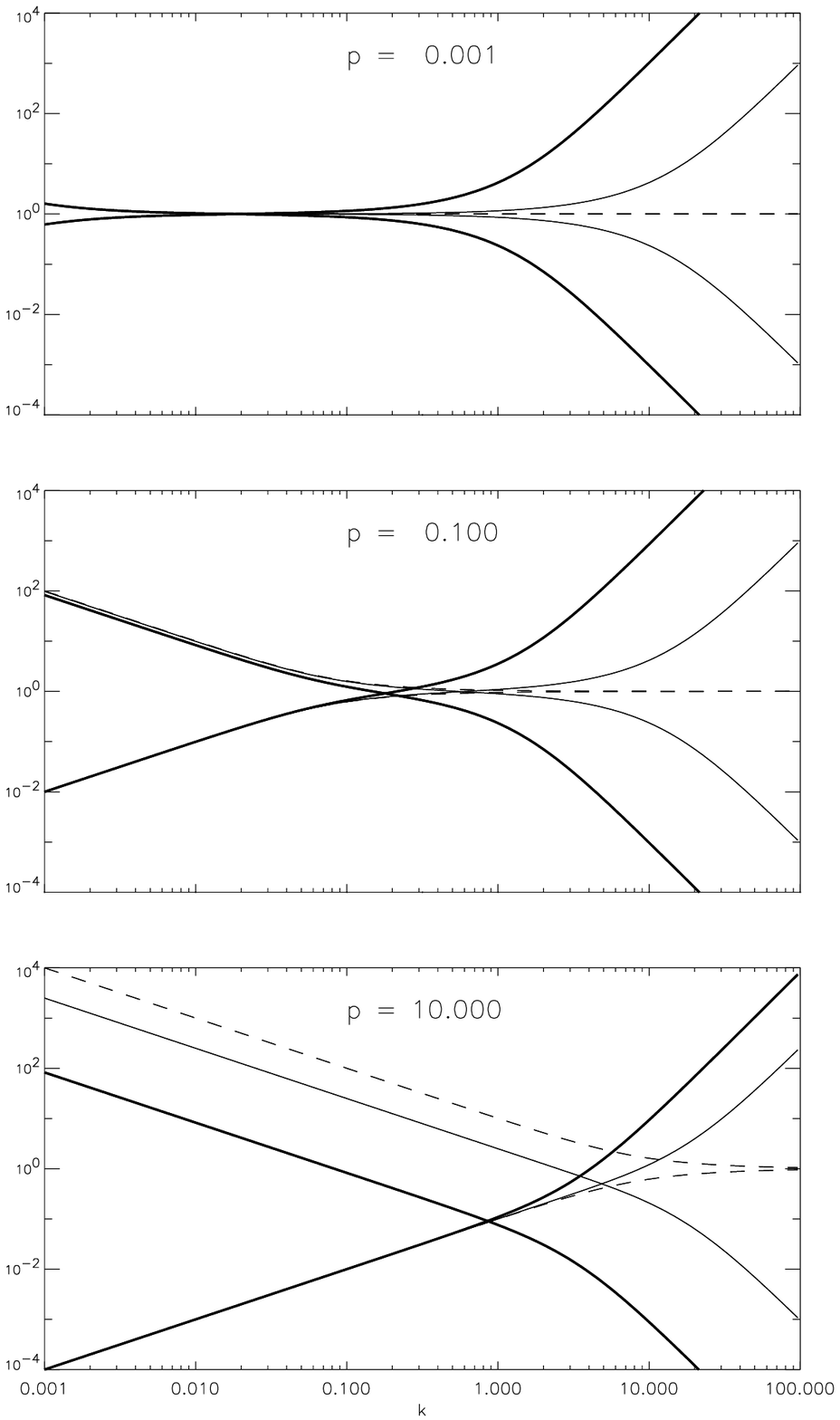}
\caption{The trace of $\alpha_{ij}$ vs. k in arbitrary units (see equation (\ref{alphatensor2}) for the 
two positive branches of the dispersion relationship. Each frame corresponds to a different value of 
the dimensionless rotation speed $p$ (indicated). The dotted curves correspond to $\epsilon = 0$, the 
thin full curves to $\epsilon = 0.1$, and the thick full curves  to $\epsilon = 1$.
\label{alphak}}
\end{figure}

\section{DISCUSSION}

We have shown that the Hall effect in conjunction with fluid rotation can
generate helicity at small scales (i.e. produce small scale helical motions) 
leading to a net $\alpha$-effect 
through the agency of the shear/ion-cyclotron normal mode of the plasma. 
This finding can be of considerable importance to the existence of
large scale dynamo action in a variety of astrophysical objects.
Our results are quite
 consistent with previous results obtained in the study of instabilities 
in accretion disks. \citet{Wardle}, \citet{Balbus}, and \citet{Sano} showed 
that the magneto-rotational instability can be either enhanced or quenched 
by the Hall effect depending on the orientation of $ \vOmega $ and $ \vB_0 $, 
which is just a manifestation of the handedness introduced by the Hall 
effect. In a future work the detailed investigation of this mode of dynamo 
action will be carried out through  direct numerical simulations.

\acknowledgements
The authors are grateful to Dr. A. Pouquet for very fruitful and enlightening 
comments.
Research of SMM  was supported by US DOE contract DE-FG03-96ER-54366
Research of DOG and PDM has been funded by grant X209/01 from the 
University of Buenos Aires. PDM is a fellow of CONICET, and DOG is a 
member of the Carrera del Investigador Cient\'{\i}fico of CONICET.


\begin{thebibliography}{} 

\bibitem[Balbus \& Terquem (2001)]{Balbus} Balbus, S.A. \& Terquem, C. 2001,
         \apj, 552, 235

\bibitem[Blackman \& Field(1999)]{Blackman} Blackman, E.G. \& Field, G.B.
         1999, \apj, 521, 597

\bibitem[Brandenburg(2001)]{Brandenburg} Brandenburg, A. 2001, \apj, 550,
         824

\bibitem[Cheng et al.(2003)]{Eyink} Chen, Q., Chen, S., \& Eyink, G. 2003,
      Phys. Fluids, 15, 361
                                                                                              
\bibitem[Ding et al.(2004)]{Ding} Ding, W.X. and 8 co-authors 2004, \prl, 93, 045002

\bibitem[Geppert \& Rheinhardt(2002)]{Geppert} Geppert, U, \& Rheinhardt, M. 2002, 
        Astron. \& Astrophys., 392, 1015 

\bibitem[G\'omez \& Mininni(2004)]{Gom2004} G\'omez, D. \& Mininni, P. 2004,
    Phys. A, in press.
                                                                                              
\bibitem[Gruzinov \& Diamond(1994)]{Gruzinov} Gruzinov, A. \& Diamond, P.H.
         1994, \prl, 72, 1651

\bibitem[Helmis(1968)]{Helmis1} Helmis, G. 1968, Mber. Dtsch. Akad. Wiss.
         Berlin, 10, 280
                                                                                                
\bibitem[Helmis (1971)]{Helmis2} Helmis, G. 1971, Beitr. Plasma Physik, 11
         417

\bibitem[Ji(1999)]{Hantao} Ji, H. 1999, \prl, 83, 3198

\bibitem[Krause \& R\"adler(1980)]{Krause} Krause, F. \& R\"adler, K.-H.,
         1980, Mean-Field Magnetohydrodynamics and Dynamo Theory, (GDR:
         Pergamon Press)

\bibitem[Meneguzzi, Frisch, \& Pouquet(1981)]{Meneguzzi} Meneguzzi, M., 
         Frisch, U., \& Pouquet, A. 1981, \prl, 47, 1060

\bibitem[Mininni, G\'omez, \& Mahajan(2002a)]{MGM1} Mininni, P.D., 
         G\'omez, D.O., \& Mahajan, S.M. 2002, \apj, 567, L81

\bibitem[Mininni, G\'omez, \& Mahajan(2003a)]{MGM2} Mininni, P.D., 
         G\'omez, D.O., \& Mahajan, S.M. 2003, \apj, 584, 1120

\bibitem[Mininni, G\'omez, \& Mahajan(2003b)]{MGM3} Mininni, P.D., 
         G\'omez, D.O., \& Mahajan, S.M. 2003, \apj, 587, 472

\bibitem[Moffatt(1970a)]{Moffat1} Moffatt, H.K. 1970, J. Fluid. Mech., 41, 435

\bibitem[Moffatt(1970b)]{Moffat2} Moffatt, H.K. 1970, J. Fluid. Mech., 44, 705

\bibitem[Moffatt(1972)]{Moffat3} Moffatt, H.K. 1972, J. Fluid. Mech., 53, 385

\bibitem[Moffatt(1978)]{Moffat4} Moffatt, H.K. 1978, Magnetic field generation 
         in electrically conducting fluids (Cambridge: Cambridge University 
         Press)

\bibitem[Pouquet, Frisch, \& L\'eorat(1976)]{Pouquet} Pouquet, A., Frisch, U., 
         \& L\'eorat, J. 1976, J. Fluid Mech., 77, 321

\bibitem[R\"adler(1969)]{Radler69} R\"adler, K.-H. 1969, Monatsber. Dt. Akad. Wiss., 11, 194 

\bibitem[Sano \& Stone(2002)]{Sano} Sano, T. \& Stone, J.M. 2002, \apj,
         570, 314.

\bibitem[Spruit(2002)]{Spruit} Spruit, H.C. 2002, Astron. \& Astrophys. 381, 932

\bibitem[Urpin(2002)]{Urpin} Urpin, V. 2002, \pre 65, 026301

\bibitem[Wardle(1999)]{Wardle} Wardle, M. 1999, \mnras, 307, 849

\end{thebibliography}
\end{document}